\documentstyle [12pt] {article}
\makeatletter

\@addtoreset{equation}{section}
\ifcase \@ptsize
\or % mods for 11 pt
\or % mods for 12 pt
\fi
\advance\textheight by \topskip

\ifcase \@ptsize
\or % mods for 11 pt
\or % mods for 12 pt
\fi

\def\@citex[#1]#2{%
\if@filesw \immediate \write \@auxout {\string \citation {#2}}\fi
\@tempcntb\m@ne \let\@h@ld\relax \def\@citea{}%
\@cite{%
  \@for \@citeb:=#2\do {%
    \@ifundefined {b@\@citeb}%
      {\@h@ld\@citea\@tempcntb\m@ne{\bf ?}%
      \@warning {Citation `\@citeb ' on page \thepage \space undefined}}%
      {\@tempcnta\@tempcntb \advance\@tempcnta\@ne%
      \@tempcntb\number\csname b@\@citeb \endcsname \relax%
      \ifnum\@tempcnta=\@tempcntb %   Number follows previous--hold on to it
        \ifx\@h@ld\relax%
          \edef \@h@ld{\@citea\csname b@\@citeb\endcsname}%
        \else%
          \edef\@h@ld{\ifmmode{-}\else--\fi\csname b@\@citeb\endcsname}%
        \fi%
      \else%   %  non-successor--dump what's held and do this one
        \@h@ld\@citea\csname b@\@citeb \endcsname%
        \let\@h@ld\relax%
      \fi}%
    \def\@citea{,\penalty\@highpenalty\,}%
  }\@h@ld%
}{#1}}

\makeatother
\begin{document}
\hfuzz=100pt
%\rightmargin -2.75cm
%\textheight 23.0cm
%\topmargin -0.5in
%\baselineskip 16pt
%\parskip 18pt
%\parindent 30pt
%\def\mc{\,\raise -2.truept\hbox{\rlap{\hbox{$\sim$}}\raise5.truept
%\hbox{$<$}\ }}
%\def\Mc{\,\raise -2.truept\hbox{\rlap{\hbox{$\sim$}}\raise5.truept
%\hbox{$>$}\ }}%
%
%
%%%%%%%%%%%%%%%%%%%%%%%%%%%%%%%%%%%%%%%%%%%%%%%%%%%%%%%%%%%%%%%
%
%    List of  the     commands
%%%%%%%%%%%%%%%%%%%%%%%%%%%%%%%%%%%%%%%%%%%%%%%%%%%%%%%%%%%%%%%
%
\newcommand{\be}{\begin{equation}}
\newcommand{\ee}{\end{equation}}
\newcommand{\bea}{\begin{eqnarray}}
\newcommand{\eea}{\end{eqnarray}}
\begin{titlepage}
\makeatletter
\def \thefootnote {\fnsymbol {footnote}} \def \@makefnmark {
\hbox to 0pt{$^{\@thefnmark }$\hss }}
\makeatother
\begin{flushright}
BONN-HE-92-33\\
November, 1992
\end{flushright}
\vspace{2cm}
\begin{center}
{ \large \bf Discrete Symmetry, Non-Commutative Geometry and Gravity }\\
\vspace{2cm}
{\large\bf Noureddine Mohammedi} \footnote
{Work supported by the Alexander von Humboldt-Stiftung.}
\footnote{e-mail: nouri@avzw02.physik.uni-bonn.de}
\\
\vspace{.5cm}
\large Physikalisches Institut\\
der Universit\"at Bonn\\
Nussallee 12\\ D-5300 Bonn 1, Germany\\

\baselineskip 18pt
\vspace{.2in}
\vspace{1cm}
{\large\bf Abstract}
\end{center}
We describe the geomety of a set of scalar fields coupled to
gravity. We consider the formalism of a differential $Z_2$-graded
algebra of $2\times 2$ matrices whose elements
are differential forms on space-time. The connection and the
vierbeins
are extended to incorporate additional scalar and vector fields.
The resulting action describes two universes coupled in a
non-minimal way to a set of scalar fields. This picture is
slightly different from the description of general relativity
in the framework of non-commutative geomety.
\\

\setcounter {footnote}{0}
\end{titlepage}
\baselineskip 20pt
\section{Introduction}
Although the standard model of electroweak interactions [1] has
successfully passed all the experimental tests at the presently
available energies, we still lack a convincing understanding
of the phenomena of spontaneous symmetry breaking. The appearance
of the Higgs fields, their self interactions and their Yukawa
couplings to the other fields in the standard model is quite
artificial. Many attempts, however, have been made to give the
Higgs sector a geometrical origin. These attempts are usually in
the form of some Kaluza-Klein theories [2] or compactified string
models. In these descriptions of the Higgs fields, space-time is
taken to be locally written as $M_4\times F$. When $M_4$ and
$F$ are smooth manifolds, one needs only the usual mathematical
tools of differential geometry. However, when the internal space
$F$ is a discrete set of points, non-commutative geometry does
apply.
\par
Recently, a geometrical picture unifying the gauge fields and the
Higgs fields of the standard $SU(2)\times U(1)$ model was put
forward by A. Connes [3,4]. The proposed space-time is a product of
a continuous Euclidean manifold and a discrete space consisting
of two points. The vector potential defined on this space has
the usual $SU(2)$ gauge fields along the continuous directions
and the $U(1)$ component in the form of a Higgs field along
the discrete direction [5-7].
Furthermore, by enlarging the discrete
space to three points, one obtains other models of particle
physics such as grand unification models [8].
\par
There is however another approach to geometrically describe the
Higgs sector and spontaneous symmetry breaking without using the
machinery of non-commutative geometry. This method is
mathematically simpler and uses the algebra of $2\times 2$ matrices
whose entries are functions or $p$-forms on space-time [9]. The
basic mathematical object here is a ``generalized connection'' whose
diagonal elements give the Yang-Mills fields and whose off-diagonal
elements characterize the Higgs fields (see also refs.[10-16] for
related topics on discrete symmetry and non-commutative
geometry).
\par
We will, in this note, apply this last approach to the theory of
gravity. Our fundamental mathematical objects are $2\times 2$
``generalized spin connection'' together with a $2\times 2$
``generalized vierbeins''. We then construct out of these objects
a gauge invariant action and obtain, in the most general case,
a theory describing two distinct universes (the left and right
universes). These two universes, however, are coupled to each
other in a non-trivial way through the presence of an action
for a set of scalar fields.
We note here that in the framework of non-commutative geometry one
obtains a model of only one single scalar field coupled in
a minimal manner to Einstein-Hilbert gravity [17].
As it is well-known, scalar theories
coupled to gravity are of crucial importance in cosmology and
account, for instance, for the inflation of the universe. Hence
giving a geometrical origin to these theories might be relevant
to cosmology.
\par
We start this note by reviewing the mathematical notions used to
describe the algebra of $2\times 2$ matrices whose elements are
differential forms. The setting for analysing gravity in the context
of this algebra is then presented and the gauge invariant action
is proposed.
\section{An Algebra of $2\times 2$ Matrices}
%\setcounter{equation}{0}
%%%%%%%%%%%%
We would like to construct in this section a differential $Z_2$-graded
algebra of $2\times 2$ matrices whose elements
are functions or $p$-forms. The most general element in this
algebra is written as
\footnote{In this section, we will present only the necessary
mathematical definitions and refer the reader to ref.[9] for more
details.}
\be
X=\left( \begin{array}{cc}
A&C\\
D&B
\end{array}\right)\,\,\,,
\ee
where $A$, $B$, $C$, and $D$ can be complex numbers, functions or
differential forms. This algebra is also going to be constructed as
the tensor product of two graded associative and differential
algebras. The first consists of the algebra of $2\times 2$
complex matrices, where the $Z_2$-grading is defined by associating
a degree $0$ to diagonal matrices (even) and $-1$ to the
off-diagonal ones (odd). The second differential graded algebra
is the algebra of differential forms (the algebra for the addition
and wedge product of forms).
\par
Let us denote by $\odot$ the product which defines this
$Z_2$-graded associative algebra. Let us also denote by
$\partial A$, $\partial B$, $\partial C$ and $\partial D$ the
degrees
of the differential forms $A$, $B$, $C$ and $D$ respectively.
The associative product of two elements in this algebra,
$X$ and $X'$, is defined as [9]
\bea
&\left( \begin{array}{cc}
A&C\\
D&B
\end{array}\right)
\odot
\left( \begin{array}{cc}
A'&C'\\
D'&B'
\end{array}\right)&\nonumber\\
&=
\left( \begin{array}{cc}
A\wedge A'+(-1)^{\partial C}C\wedge D'&
C\wedge B'+(-1)^{\partial A}A\wedge C'\\
D\wedge A'+(-1)^{\partial B}B\wedge D'&
B\wedge B'+(-1)^{\partial D}D\wedge C'\end{array}\right)\,\,\,.&
\eea
\par
The next step would be to define a differential operator $\hat d$
acting on elements like $X$,
satisfying graded Leibniz rule and being nilpotent $\hat {d}^2\,
=\,0$. This would be a
generalization of the notion of the usual exterior derivative $d$
in ordinary differential geometry. The action of $\hat{d}$ on
$X$ is given by [9]
\be
\hat{d}X=
\left( \begin{array}{cc}
dA+C+D&-dC-(A-B)\\
-dD+(A-B)&dB+C+D
\end{array}\right)\,\,\,.
\ee
Notice that $\hat{d}X\,\neq\,0$ even when $A$, $B$, $C$ and $D$ are
just complex numbers. It is also easy to see that $\hat{d}^2X\,=\,
0$. Furthermore, $\hat{d}$ obeys graded Leibniz rule
\be
\hat{d}\left(X\odot X'\right)=
\hat{d}X\odot X'+\left(-1\right)^{\partial X}X\odot \hat{d}X'
\,\,\,.
\ee
Here $\partial X$ is the total $Z_2$-grading of $X$. That is,
$\partial X$ is the sum of the $Z_2$-grading of $X$ as a matrix
(diagonal or off-diagonal) and as a differential form (even or
odd degree). To compute the second term of the right-hand side one
has, therefore, to write $X$ as a sum of four matrices each having
only one non-zero element.
The previous mathematical ingredients are
just what we need to construct
the gravitational theory that we have in mind. We will work with a
``generalized spin connection'', $\Omega^a_{\,\,\,b}$,  and a
``generalized vierbein'', $E^a$, in the form of $2\times 2$ matrices
\be
\Omega^a_{\,\,\,b}=
\left(\begin{array}{cc}
\omega^a_{\,\,\,b}&\varphi^a_{\,\,\,b}\\
\widetilde\varphi^a_{\,\,\,b}&\widetilde\omega^a_{\,\,\,b}
\end{array}\right)\,\,\,,\,\,\,
E^a=
\left(\begin{array}{cc}
e^a&s^a\\
\widetilde s^a&\widetilde e^a
\end{array}\right)\,\,\,.
\ee
Here $\omega^a_{\,\,\,b}$, $\widetilde\omega^a_{\,\,\,b}$, $e^a$,
$\widetilde e^a$ are all real one-forms and we have
$\omega^a_{\,\,\,b}=\omega^a_{\mu b}dx^\mu$,
$\widetilde\omega^a_{\,\,\,b}=\widetilde\omega^a_{\mu b}dx^\mu$,
$e^a=e^a_\mu dx^\mu$,
$\widetilde e^a=\widetilde e^a_\mu dx^\mu$, while
$\varphi^a_{\,\,\,b}$, $\widetilde\varphi^a_{\,\,\,b}$, $s^a$,
$\widetilde s^a$ are real functions.
\par
The generalized curvature is defined as
\be
{\cal {R}}^a_{\,\,\,b}=\hat
d\Omega^a_{\,\,\,b}+\Omega^a_{\,\,\,c}\odot\Omega^c_{\,\,\,b}
\,\,\,.
\ee

In analogy with ordinary differential geometry, we define the
generalized torsion to be given by
\be
{\cal {T}}^a=\hat d E^a+\Omega^a_{\,\,\,b}\odot E^b\,\,\,.
\ee
The last two equations are the generalization of Cartan's structure
equations in differential geometry. Furthermore, by acting on both
sides of (2.6) with $\hat{d}$,
we find the generalized Bianchi identities
\be
\hat d{\cal {R}}^a_{\,\,\,b}+\Omega^a_{\,\,\,c}\odot
{\cal {R}}^a_{\,\,\,c}-{\cal {R}}^a_{\,\,\,c}\odot\Omega^c_{\,\,\,b}
\equiv {\cal {D}}{\cal {R}}^a_{\,\,\,b}=0\,\,\,.
\ee
The action of $\hat{d}$ on (2.7) gives the following consistency
conditions
\be
\hat d{\cal {T}}^a+\Omega^a_{\,\,\,c}\odot{\cal {T}}^c=
{\cal {R}}^a_{\,\,\,c}\odot E^c\,\,\,.
\ee
In deriving the above two consistency equations, we have used the
fact that the total $Z_2$-grading of $\Omega^a_{\,\,\,b}$ is $+1$.
This is because the diagonal elements of $\Omega^a_{\,\,\,b}$ are
one-forms and its off-diagonal terms are zero-forms.
\par
The matrix for the curvature is found to be
\be
{\cal {R}}^a_{\,\,\,b}=
\left(\begin{array}{cc}
({\cal {R}}_{11})^a_b&({\cal {R}}_{12})^a_b\\
({\cal {R}}_{21})^a_b&({\cal {R}}_{22})^a_b\end{array}\right)\,\,\,,
\ee
with
\bea
({\cal {R}}_{11})^a_b&=&R^a_{\,\,\,b}+\varphi^a_{\,\,\,b}+
\widetilde\varphi^a_{\,\,\,b}+\varphi^a_{\,\,\,c}\widetilde\varphi
^c_{\,\,\,b}\nonumber\\
({\cal {R}}_{12})^a_b&=&-\nabla\varphi^a_{\,\,\,b}-\left(\omega^a_{\,\,\,b}
-\widetilde\omega^a_{\,\,\,b}\right)\nonumber\\
({\cal {R}}_{21})^a_b&=&-\nabla\widetilde\varphi^a_{\,\,\,b}+
\left(\omega^a_{\,\,\,b}
-\widetilde\omega^a_{\,\,\,b}\right)\nonumber\\
({\cal {R}}_{22})^a_b&=&\widetilde R^a_{\,\,\,b}+\varphi^a_{\,\,\,b}+
\widetilde\varphi^a_{\,\,\,b}+\widetilde\varphi
^a_{\,\,\,c}\varphi^c_{\,\,\,b}\,\,\,,
\eea
where
\bea
R^a_{\,\,\,b}&=&d\omega^a_{\,\,\,b}+\omega^a_{\,\,\,c}\wedge
\omega^c
_{\,\,\,b}={1\over 2}R^a_{\,\,\,b\mu\nu}dx^\mu\wedge dx^\nu
\nonumber\\
\widetilde R^a_{\,\,\,b}&=&d\widetilde\omega^a_{\,\,\,b}+
\widetilde\omega^a_{\,\,\,c}\wedge\widetilde\omega^c
_{\,\,\,b}={1\over 2}\widetilde R^a_{\,\,\,b\mu\nu}
dx^\mu\wedge dx^\nu
\nonumber\\
\nabla\varphi^a_{\,\,\,b}&=&d\varphi^a_{\,\,\,b}-
\varphi^a_{\,\,\,c}
\widetilde\omega^c_{\,\,\,b}+\omega^a_{\,\,\,c}\varphi^c_{\,\,\,b}
\nonumber\\
\nabla\widetilde\varphi^a_{\,\,\,b}&=&d\widetilde
\varphi^a_{\,\,\,b}-\widetilde\varphi^a_{\,\,\,c}
\omega^c_{\,\,\,b}+\widetilde\omega^a_{\,\,\,c}
\widetilde\varphi^c_{\,\,\,b}\,\,\,.
\eea
Similarly, the torsion matrix is written as
\be
{\cal {T}}^a=\left(\begin{array}{cc}
{\cal {T}}^a_{11}&{\cal {T}}^a_{12}\\
{\cal {T}}^a_{21}&{\cal {T}}^a_{22}\end{array}\right)\,\,\,,
\ee
where
\bea
{\cal {T}}^a_{11}&=& de^a+\omega^a_{\,\,\,b}\wedge e^b+s^a+
\widetilde s^a+\varphi^a_{\,\,\,b}\widetilde s^b\nonumber\\
{\cal {T}}^a_{12}&=& -\nabla s^a-(e^a-\widetilde e^a) +\varphi^a_{\,\,\,b}
\widetilde e^b\nonumber\\
{\cal {T}}^a_{21}&=& -\nabla \widetilde s^a+(e^a-\widetilde e^a)
+\widetilde\varphi^a_{\,\,\,b}e^b\,\,\,\nonumber\\
{\cal {T}}^a_{22}&=& d\widetilde e^a+\widetilde\omega^a_{\,\,\,b}
\wedge \widetilde e^b+s^a+
\widetilde s^a+\widetilde\varphi^a_{\,\,\,b} s^b\,\,\,,
\eea
with
\bea
\nabla s^a&=&ds^a+\omega^a_{\,\,\,b}s^b\nonumber\\
\nabla \widetilde s^a&=&d\widetilde
s^a+\widetilde\omega^a_{\,\,\,b}\widetilde s^b\,\,\,.
\eea
\par
We would like now to to examine the issue of gauge transformations.
In analogy with differential geometry, we consider a generalized
orthogonal rotation of the generalized orthonormal frame
\be
E^a \longrightarrow E'^a={\cal {H}}^a_{\,\,\,b}\odot E^b\,\,\,.
\ee
here ${\cal {H}}^a_{\,\,\,b}$ is a $2\times 2$ matrix whose entries are
all zero-forms. This takes the general form
\be
{\cal {H}}^a_{\,\,\,b}=\left(\begin{array}{cc}
h^a_{\,\,\,b}&f^a_{\,\,\,b}\\
\widetilde f^a_{\,\,\,b}&\widetilde h^a_{\,\,\,b}
\end{array}\right)
\ee
We define in what follows the generalized Cartesian flat metric,
$\Sigma_{ab}$, and the generalized Kronecker delta-function,
$\Upsilon^a_{\,\,\,b}$, as
\be
\Sigma_{ab}=\left ( \begin{array}{cc}
\eta_{ab}&0\\
0&\eta_{ab}\end{array}\right)\,\,\,,\,\,\,
\Upsilon^a_{\,\,\,b}=\left(\begin{array}{cc}
\delta^a_{\,\,\,b}&0\\
0&\delta^a_{\,\,\,b}\end{array}\right)\,\,\,.
\ee
The matrix ${\cal {H}}^a_{\,\,\,b}$ satisfies
\bea
\Sigma_{cd}&=&\Sigma_{ab}\odot {\cal {H}}^a_{\,\,\,c}\odot
{\cal {H}}^b_{\,\,\,d}\nonumber\\
\Upsilon^a_{\,\,\,b}&=&{\cal {H}}^a_{\,\,\,c}\odot
\left({\cal {H}}^{-1}\right)^c_{b}\nonumber\\
\left(\hat {d}{\cal {H}}^a_{\,\,\,c}\right)\odot
\left({\cal {H}}^{-1}\right)^c_{b}&=&
\left[\left({\cal {H}}^a_{\,\,\,c}\right)_o-\left({\cal
{H}}^a_{\,\,\,c}\right)_e
\right]\odot \left(\hat{d}{\cal {H}}^{-1}\right)^c_b\,\,\,,
\eea
where the subscripts $e$ and $o$ stand, respectively, for the
diagonal (even) and off-diagonal (odd) parts of $2\times 2$
matrices.
\par
We require the torsion to transform like
\be
{\cal {T}}'^a=\hat d E'^a+\Omega '^a_{\,\,\,b}\odot E'^b
={\cal {H}}^a_{\,\,\,b}\odot {\cal {T}}^b\,\,\,,
\ee
and the connection as
\be
\Omega'^a_{\,\,\,b}={\cal {H}}^a_{\,\,\,c}\odot\Omega^c_{\,\,\,d}
\odot\left({\cal {H}}^{-1}\right)^d_b+{\cal {H}}^a_{\,\,\,c}
\odot\left(\hat d{\cal {H}}^{-1}\right)^c_b\,\,\,.
\ee
By computing ${\cal {T}}'^a\,=\,{\cal {H}}^a_{\,\,\,b}\odot {\cal {T}}^b$
and looking at the resulting transformations for the elements
of $\Omega^a_{\,\,\,b}$ and $E^a$, we deduce that
${\cal {H}}^a_{\,\,\,b}$ must be such that $f^a_{\,\,\,b}\,=\,
\widetilde f^a_{\,\,\,b}\,=\,0$ and $h^a_{\,\,\,b}\,=\,
\widetilde h^a_{\,\,\,b}$. Therefore, $\left({\cal {H}}^{-1}\right)^a_b
$ has $\left(h^{-1}\right)^a_b$ along the diagonal and $0$ along
the off-diagonal. Consequently, the curvature transforms as
expected
\be
{\cal {R}}'^a_{\,\,\,b}=\hat d \Omega'^a_{\,\,\,b}+\Omega
'^a_{\,\,\,c}\odot \Omega'^c_{\,\,\,b}
={\cal {H}}^a_{\,\,\,c}\odot {\cal {R}}^c_{\,\,\,d}\odot\left(
{\cal {H}}^{-1}\right)^d_b\,\,\,.
\ee
The elements of $\Omega^a_{\,\,\,b}$ and $E^a$ transform in the
following way
\bea
e'^a&=&h^a_{\,\,\,b}s^b\,\,,\,\, \widetilde e'^a=h^a_{\,\,\,b}
\widetilde e^b\,\,,\,\,
s'^a=h^a_{\,\,\,b}s^b\,\,,\,\, \widetilde s'^a=h^a_{\,\,\,b}
\widetilde s^b\nonumber\\
\omega'^a_{\,\,\,b}&=&h^a_{\,\,\,c}\omega^c_{\,\,\,d}(h^{-1})^d_b
+h^a_{\,\,\,c}(dh^{-1})^c_b
\,\,,\,\,
\widetilde\omega'^a_{\,\,\,b}=h^a_{\,\,\,c}
\widetilde\omega^c_{\,\,\,d}(h^{-1})^d_b
+h^a_{\,\,\,c}(dh^{-1})^c_b\nonumber\\
\varphi'^a_{\,\,\,b}&=&h^a_{\,\,\,c}\varphi^c_{\,\,\,d}(h^{-1})^d_b
\,\,,\,\,
\widetilde\varphi'^a_{\,\,\,b}=h^a_{\,\,\,c}
\widetilde\varphi^c_{\,\,\,d}(h^{-1})^d_b\,\,\,.
\eea
Notice that $s^a$ and $\widetilde s^a$ transform like vectors
while $\varphi^a_{\,\,\,b}$ and $\widetilde\varphi^a_{\,\,\,b}$
transform like scalars.

\section{The Action}

We turn now to the construction of an action which is gauge
invariant under the above gauge transformations.
Using the scalar product of differential forms, where the
space-time metric is $g_{\mu\nu}$, and the trace on the space of
$2\times 2$ matrices, we
define the following gauge invariant Lagrangian
\bea
S &=& \int Tr\left [\left(E^a\odot E^b\right)\odot
\ast\left(\Sigma_{bc}\odot{\cal {R}}^c_{\,\,\,a}\right)\right]
\nonumber\\
&=&\int\left[\left (E^a\odot E^b\right)_{11}\wedge
\ast\left(\Sigma_{bc}\odot{\cal {R}}^c_{\,\,\,a}\right)_{11}
-\left(E^a\odot E^b\right)_{12}\wedge
\ast\left(\Sigma_{bc}\odot{\cal {R}}^c_{\,\,\,a}\right)_{21}
\right.
\nonumber\\
&-&\left(E^a\odot E^b\right)_{21}\wedge
\ast\left(\Sigma_{bc}\odot{\cal {R}}^c_{\,\,\,a}\right)_{12}
\nonumber\\
&+&\left.\left(E^a\odot E^b\right)_{22}\wedge
\ast\left(\Sigma_{bc}\odot{\cal {R}}^c_{\,\,\,a}\right)_{22}
\right]\,\,\,.
\eea
Here the integration is over a $n$-dimensional space and $\ast$
is the usual Hodge star which acts on the individual elements of
the matix $(\Sigma_{ac}\odot{\cal{R}}^c_{\,\,\,b})$.
An explicit computation gives
\bea
S &=&\int d^nx\sqrt{g}\left[\left(e^a_\mu e^b_\nu\eta_{bc} R^c_{\,\,\,a\rho
\sigma} +
\widetilde e^a_\mu\widetilde  e^b_\nu\ \eta_{bc}
 R^c_{\,\,\,a\rho
\sigma}\right)g^{\mu\rho}g^{\nu\sigma}\right.\nonumber\\
&+&\left(s^a\widetilde
e^b_\mu - e^a_\mu s^b\right)\eta_{bc}
 D_\nu\widetilde \varphi^c_{\,\,\,a}g^{\mu\nu}
+\left(\widetilde s^a
e^b_\mu -\widetilde e^a_\mu \widetilde s^b\right)\eta_{bc}
D_\nu\varphi^c_{\,\,\,a}g^{\mu\nu}\nonumber\\
&+&s^a\widetilde s^b\eta_{bc}\left(
\varphi^c_{\,\,\,a}+\widetilde\varphi^c_{\,\,\,a}
+\varphi^c_{\,\,\,d}\widetilde\varphi^d_{\,\,\,a}\right)
\nonumber\\
&+&\left.\widetilde s^as^b\eta_{bc}\left(
\varphi^c_{\,\,\,a}+\widetilde\varphi^c_{\,\,\,a}
+\widetilde\varphi^c_{\,\,\,d}\varphi^d_{\,\,\,a}\right)\right]
\,\,\,,
\eea
where
\bea
D_{\mu}\varphi^a_{\,\,\,b}&=&\nabla_{\mu}\varphi^a_{\,\,\,b}+
\left(\omega^a_{\mu b}-\widetilde\omega^a_{\mu b}\right)
\nonumber\\
D_{\mu}\widetilde\varphi^a_{\,\,\,b}&=&\nabla_{\mu}
\widetilde\varphi^a_{\,\,\,b}-
\left(\omega^a_{\mu b}-\widetilde\omega^a_{\mu b}\right)\,\,\,.
\eea
\par
At this stage all the fields entering in the construction of the
above Lagrangian are independent of each other. One can, therefore,
simply take $S$ as a starting point and eliminate all the
non-propagating fields by their equations of motion.
Indeed, the equations of motion for $s^a$ and $\widetilde s^a$
can be easily solved and we find
\bea
s^a&=&\left(M^{-1}\right)^{ab}\left(\widetilde e^d_\mu\eta_{bc}
D_\nu\varphi^c_{\,\,\,d}-e^d_\mu\eta_{dc}D_\nu\varphi^c_{\,\,\,b}
\right)g^{\mu\nu}\nonumber\\
\widetilde s^a&=&\left(N^{-1}\right)^{ab}\left( e^d_\mu\eta_{bc}
D_\nu\widetilde\varphi^c_{\,\,\,d}-e^d_\mu\eta_{dc}D_\nu
\widetilde\varphi^c_{\,\,\,b}
\right)g^{\mu\nu}\,\,\,,
\eea
where
\bea
M_{ab}&=&\eta_{ac}\left(
\varphi^c_{\,\,\,b}+\widetilde\varphi^c_{\,\,\,b}
+\varphi^c_{\,\,\,d}\widetilde\varphi^d_{\,\,\,b}\right)
+\eta_{bc}\left(
\varphi^c_{\,\,\,a}+\widetilde\varphi^c_{\,\,\,a}
+\widetilde\varphi^c_{\,\,\,d}\varphi^d_{\,\,\,a}\right)
\nonumber\\
N_{ab}&=&\eta_{ac}\left(
\varphi^c_{\,\,\,b}+\widetilde\varphi^c_{\,\,\,b}
+\widetilde\varphi^c_{\,\,\,d}\varphi^d_{\,\,\,b}\right)
+\eta_{bc}\left(
\varphi^c_{\,\,\,a}+\widetilde\varphi^c_{\,\,\,a}
+\varphi^c_{\,\,\,d}\widetilde\varphi^d_{\,\,\,a}\right)\,\,\,.
\eea
By substituting for $s^a$ and $\widetilde s^a$ in the above action
we get
\bea
S &=&\int d^nx\sqrt{g}\left[\left(e^a_\mu e^b_\nu\eta_{bc} R^c_{\,\,\,a\rho
\sigma} +
\widetilde e^a_\mu\widetilde  e^b_\nu\eta_{bc}
 R^c_{\,\,\,a\rho
\sigma}\right)g^{\mu\rho}g^{\nu\sigma}\right.\nonumber\\
&+&\left.{\cal{F}}^{da,\sigma\nu}_{cs}D_\sigma\varphi^c_{\,\,\,d}
D_\nu\widetilde\varphi^s_{\,\,\,a}\right]\,\,\,,
\eea
where the quantity ${\cal{F}}^{da,\sigma\nu}_{cs}$ is given by
\bea
{\cal{F}}^{da,\sigma\nu}_{cs}&=&g^{\mu\nu}g^{\rho\sigma}
\eta_{rc}\eta_{bs}\left[\left(M^{-1}\right)^{ar}\widetilde e^d_\rho
\widetilde e^b_\mu
+\left(M^{-1}\right)^{bd} e^a_\mu
 e^r_\rho\right.\nonumber\\
&-&\left.\left(M^{-1}\right)^{ad}e^r_\rho
\widetilde e^b_\mu-\left(M^{-1}\right)^{br} e^a_\mu
\widetilde e^d_\rho\right]\nonumber\\
&+&g^{\rho\nu}g^{\mu\sigma}
\eta_{rs}\eta_{bc}\left[\left(N^{-1}\right)^{dr}\ e^a_\rho
 e^b_\mu
+\left(N^{-1}\right)^{ba}\widetilde e^d_\mu
 \widetilde e^r_\rho\right.\nonumber\\
&-&\left.\left(N^{-1}\right)^{da}\widetilde e^r_\rho
 e^b_\mu-\left(N^{-1}\right)^{br} \widetilde e^d_\mu
 e^a_\rho\right]\nonumber\\
&+&g^{\mu\sigma}g^{\rho\nu}\eta_{rc}\eta_{fs}\eta_{bl}
\left(\varphi^l_{\,\,\,t}+\widetilde\varphi^l_{\,\,\,t}
+\varphi^l_{\,\,\,k}\widetilde\varphi^k_{\,\,\,t}\right)
\left[\left(M^{-1}\right)^{tr}\left(N^{-1}\right)^{bf}
\widetilde e^d_\mu e^a_\rho\right.\nonumber\\
&-&\left(M^{-1}\right)^{tr}\left(N^{-1}\right)^{ba}
\widetilde e^d_\mu \widetilde e^f_\rho
-\left(M^{-1}\right)^{td}\left(N^{-1}\right)^{bf}
 e^r_\mu e^a_\rho\nonumber\\
&+&\left.\left(M^{-1}\right)^{td}\left(N^{-1}\right)^{ba}
e^r_\mu \widetilde e^f_\rho\right]\nonumber\\
&+&g^{\mu\nu}g^{\rho\sigma}\eta_{rs}\eta_{fc}\eta_{bl}
\left(\varphi^l_{\,\,\,t}+\widetilde\varphi^l_{\,\,\,t}
+\widetilde\varphi^l_{\,\,\,k}\varphi^k_{\,\,\,t}\right)
\left[\left(N^{-1}\right)^{tr}\left(M^{-1}\right)^{bf}
 e^a_\mu \widetilde e^d_\rho\right.\nonumber\\
&-&\left(N^{-1}\right)^{tr}\left(M^{-1}\right)^{bd}
 e^a_\mu  e^f_\rho
-\left(N^{-1}\right)^{ta}\left(M^{-1}\right)^{bf}
 \widetilde e^r_\mu \widetilde e^d_\rho\nonumber\\
&+&\left.\left(N^{-1}\right)^{ta}\left(M^{-1}\right)^{bd}
\widetilde e^r_\mu  e^f_\rho\right]\,\,\,.
\eea
The resulting action describes two distinct universes (the
``tilded'' universe and the ``untilded'' one) which are coupled
through the presence of the ${\cal{F}}^{da,\sigma\nu}_{cs}$ term.
In order to obtain Einstein-Hilbert gravity, we restrict ourselves
to the case when
\be
\widetilde e^a_\mu=e^a_\mu\,\,,\,\,\widetilde\omega_{\mu b}^a
=\omega^a_{\mu b}\,\,,\,\,g_{\mu\nu}=e^a_\mu e^b_\nu\eta_{ab}
\ee
and $\omega^a_{\mu b}$ is the Levi-Civita spin connection
satisfying the metricity condition ($\omega_{ab}\,=\,\eta_{ac}
\omega^{c}_{\,\,\,b}=-\omega_{ba}$) and the torsion-free
constraints ($de^a+\omega^a_{\,\,\,b}\wedge e^b\,=0\,$). The first
term in the action (3.6) is then the Einstein-Hilbert action while
the second term describes a non-minimal coupling of a set of
scalar fields to gravity.
\par
The other possibility in dealing with our starting action (3.2)
would
be to impose the generalized torsion-free conditions, ${\cal{T}}^a
\,=\,0$. It turns out, however, that these constraints are very
restrictive and, apart from some trivial solutions, no general
solutions were found.
In what follows we will propose another definition for the curvature
and the torsion in such a way that the torsion-free constraints
can be solved.
\par
We define the new generalized curvature as
\be
\hat{{\cal {R}}}^a_{\,\,\,b}=\hat d \Omega^a_{\,\,\,b}+\Omega
^a_{\,\,\,c}\odot \Omega^c_{\,\,\,b}-\left(\Omega
^a_{\,\,\,c}\right)_o\odot \left(\Omega^c_{\,\,\,b}\right)_o
\,\,\,,
\ee
where $\left(\Omega^a_{\,\,\,b}\right)_o$ is the off-diagonal
part of $\Omega^a_{\,\,\,b}$. It is easy to see that under
the gauge transformations ${\cal {H}}^a_{\,\,\,b}$ (for which
$f^a_{\,\,\,b}\,=\,\widetilde f^a_{\,\,\,b}\,=\,0$ and
$h^a_{\,\,\,b}\,=\,\widetilde h^a_{\,\,\,b}$) the new curvature
tensor transforms as
\be
\hat{{\cal {R}}}'^a_{\,\,\,b}={\cal {H}}^a_{\,\,\,c}\odot \hat{{\cal
{R}}}^c_{\,\,\,d}\odot\left({\cal {H}}^{-1}\right)^d_b\,\,\,.
\ee
Similarly, the new torsion is given by
\be
\hat{{\cal {T}}}'^a=\hat d E^a+\Omega
^a_{\,\,\,c}\odot E^c-\left(\Omega
^a_{\,\,\,b}\right)_o\odot \left(E^b\right)_o\,\,\,,
\ee
and it transforms as
\be
\hat{{\cal {T}}}'^a={\cal {H}}^a_{\,\,\,b}\odot \hat{{\cal {T}}}^b\,\,\,.
\ee
The elements of $\hat{{\cal {R}}}^a_{\,\,\,b}$ are simply given by the
expressions in (2.12) but without the terms $\varphi^a_{\,\,\,c}
\widetilde\varphi^c_{\,\,\,b}$ and $\widetilde\varphi^a_{\,\,\,c}
\varphi^c_{\,\,\,b}$ in $({\cal {R}}_{11})^a_b$ and $({\cal {R}}_{22})^a_b$
respectively. Also the elements of $\hat{{\cal {T}}}^a$ are given by
the expressions in (2.14) but without the terms $\varphi^a_{\,\,\,b}
\widetilde s^b$ and $\widetilde\varphi^a_{\,\,\,b}s^b$ in
$({\cal{T}}^a)_{11}$ and $({\cal{T}}^a)_{22}$ respectively.
\par
The new gauge invariant action is given by
\bea
S &=& \int Tr\left [\left(E^a\odot E^b\right)\odot
\ast\left(\Sigma_{bc}\odot\hat{{\cal {R}}}^c_{\,\,\,a}\right)\right]
\nonumber\\
&=&\int d^nx\sqrt{g}\left[\left(e^a_\mu e^b_\nu\eta_{bc} R^c_{\,\,\,a\rho
\sigma} +
\widetilde e^a_\mu\widetilde  e^b_\nu\eta_{bc}
 R^c_{\,\,\,a\rho
\sigma}\right)g^{\mu\rho}g^{\nu\sigma}\right.\nonumber\\
&+&\left[\left(s^a\widetilde
e^b_\mu - e^a_\mu s^b\right)\eta_{bc}
D_\nu\widetilde\varphi^c_{\,\,\,a}
+\left(\widetilde s^a
e^b_\mu - \widetilde e^a_\mu \widetilde s^b\right)\eta_{bc}
 D_\nu\varphi^c_{\,\,\,a}\right]g^{\mu\nu}
\nonumber\\
&+&\left.\left(s^a\widetilde s^b+\widetilde s^as^b\right)\eta_{bc}
\left(\varphi^c_{\,\,\,a}+\widetilde\varphi^c_{\,\,\,a}
\right)\right]\,\,\,.
\eea
The torsion-free constraints, $\hat{{\cal {T}}}^a\,=\,0$, are solved by
\bea
g_{\mu\nu}&=&\eta_{ab}e^a_\mu e^b_\nu\,\,,\,\,e^a=\widetilde e^a
\,\,,\,\,\omega^a_{\,\,\,b}=\widetilde\omega^a_{\,\,\,b}
\nonumber\\
\widetilde s^a&=&-s^a\,\,,\,\,\widetilde\varphi^a_{\,\,\,b}
=-\varphi^a_{\,\,\,b}=\lambda^\mu_b\nabla_\mu s^a\,\,\,,
\eea
where $\omega^a_{\,\,\,b}$ is the Levi-Civita spin connection and
 $\lambda^\mu_a$ is the inverse of $e^a_\mu$, such that
$\lambda^\mu_ae^b_\mu\,=\,\delta^b_{\,\,\,a}$,
$g^{\mu\nu}\,=\,\lambda^\mu_a\lambda^\nu_b\eta^{ab}$ and
$\eta_{ab}\,=\,g_{\mu\nu}\lambda^\mu_a\lambda^\nu_b$.
\par
Substituting these expressions in the action (3.13) and integrating by
parts, we find
\be
S=2\int d^nx\sqrt{g}\{R-\left[g^{\mu\nu}\eta_{ab}
-{1\over 2}\left(\lambda^\mu_a\lambda^\nu_b+ \lambda^\mu_b\lambda^\nu_a
\right)\right]\nabla_\mu s^a
\nabla_\nu s^b\}\,\,\,.
\ee
This is the action for a vector field coupled in a non-conventional
manner to Einstein-Hilbert gravity.
\par
To summarize, we have considered gravity in the framework of a
differential $Z_2$-graded and associative algebra of $2\times 2$
matrices with $p$-forms entries. We have given an extension of
Cartan's structure equations and constructed a gauge invariant
action. The extended torsion-free conditions are very restrictive
and would get rid of all the extra scalar fields in the theory.
Instead, we eliminate the non-dynamical fields by their equations
of motion and obtain an action describing a non-minimal coupling
of a set of scalar fields to two distinct universes. It is also
possible to modify the definitions of the extended curvature and
the extended torsion in such a way that the extended
torsion-free constraints can be solved. We obtain, in this last
case, a theory characterizing the coupling of a vector field
to Einstein-Hilbert gravity.
\vspace{0.5cm}

\paragraph{Acknowledgements:} I would like to thank Jose
Figueroa-O'Farrill
for  many useful discussions. The
financial support  from the Alexander von
Humboldt-Stiftung is also hereby acknowledged.

\end{document}